\RequirePackage{lineno}
\documentclass[twocolumn,aps,prc,showpacs,superscriptaddress,preprintnumbers,floatfix,nofootinbib]{revtex4}
\usepackage{epsfig,graphics}
\usepackage{graphicx}
\usepackage{dcolumn}
\usepackage{bm}
\usepackage{amsmath}
\usepackage[usenames]{color}
\usepackage{ulem} 

\begin{document}

\title{Equations of motion of test particles for solving the spin-dependent Boltzmann-Vlasov equation}
\author{Yin Xia}
\affiliation{Shanghai Institute of Applied Physics, Chinese Academy
of Sciences, Shanghai 201800, China}
\affiliation{University of Chinese Academy of Science, Beijing 100049, China}
\author{Jun Xu\footnote{Corresponding author: xujun@sinap.ac.cn}}
\affiliation{Shanghai Institute of Applied Physics, Chinese Academy
of Sciences, Shanghai 201800, China}
\author{Bao-An Li}
\affiliation{Department of Physics and Astronomy, Texas A$\&$M
University-Commerce, Commerce, TX 75429-3011, USA}
\affiliation{Department of Applied Physics, Xi'an Jiao Tong University, Xi'an 710049, China}
\author{Wen-Qing Shen}
\affiliation{Shanghai Institute of Applied Physics, Chinese Academy
of Sciences, Shanghai 201800, China}

\date{\today}

\begin{abstract}
A consistent derivation of the equations of motion (EOMs) of test
particles for solving the spin-dependent Boltzmann-Vlasov equation
is presented. The resulting EOMs in phase space are similar to the canonical equations in
Hamiltonian dynamics, and the EOM of spin is the same as that in
the Heisenburg picture of quantum mechanics. Considering further the quantum nature of spin and choosing the direction of total angular momentum in heavy-ion reactions as a reference of measuring nucleon spin,
the EOMs of spin-up and spin-down nucleons are given separately. The key elements affecting the spin dynamics in heavy-ion collisions are identified. The resulting EOMs provide a solid foundation for using the test-particle
approach in studying spin dynamics in heavy-ion collisions at intermediate energies.  Future comparisons of model simulations with experimental data will help constrain the poorly known in-medium nucleon spin-orbit coupling relevant for understanding properties of rare isotopes and their astrophysical impacts.
\end{abstract}

\maketitle

{\it Introduction:}~The importance of nucleon spin degree of freedom
was first recognized more than 50 years ago when Mayer and Jensen
introduced the spin-orbit interaction and used it to explain
successfully the magic numbers and shell structure of
nuclei~\cite{May49, May55}. Subsequently, the nuclear spin-orbit
interaction was found responsible for many interesting phenomena in
nuclear structure~\cite{Lal98,Mor08,Sch04,Ben09,Che95,Kra07}. It also affects some features of nuclear reactions,
such as the fusion threshold~\cite{Uma86}, the polarization measured
in terms of the analyzing power in pick-up or removal
reactions~\cite{Ich12,Sch94,Gro03,Tur06}, and the spin dependence of
nucleon collective flow~\cite{Xu13,Xia14} in heavy-ion collisions
(HICs). However, the role of nucleon spin is much less known in nuclear
reactions than structures. In HICs at intermediate
energies, a central issue is the density and isospin dependence of
the spin-orbit coupling in neutron-rich medium, see, e.g.,
Ref.~\cite{Xu15} for a recent review. It is also interesting to
mention that the study of spin-dependent structure functions of
nucleons and nuclei has been at the forefronts of nuclear and
particle physics \cite{RHICspin}. This study will be boosted by
future experiments at the proposed electron-ion collider using
polarized beams \cite{NSAC}.  In this work, we derive for the first time
equations of motion (EOMs) of nucleon test particles~\cite{Won03} for solving the
spin-dependent Boltzmann (Vlasov)-Uehling-Uhlenbeck
(BUU or VUU) equation. These EOMs provide the physics foundation for simulating spin transport for not only nucleons in heavy-ion reactions
but also electrons for understanding many interesting phenomena, such as the spin wave~\cite{Hal69,Joh84,Smi89,Ber96}, the spin-Hall effect~\cite{Sch06,Sin15,Mor15}, etc.

Considering the spin degree of freedom, the Wigner function in
phase space becomes a $2\times 2$ matrix~\cite{Conn84}. Its time
evolution is governed by the Boltzmann-Vlasov (BV) equation obtained
by a Wigner transformation of the Liouville equation for the density
matrix~\cite{Ring80,Smi89,Balb13}
\begin{eqnarray}\label{BLE}
\frac{\partial \hat{f}}{\partial t}&+&\frac{i}{\hbar}\left [ \hat{\varepsilon},\hat{f}\right]+\frac{1}{2}\left ( \frac{\partial \hat{\varepsilon}}{\partial \vec{p}}\cdot \frac{\partial \hat{f}}{\partial \vec{r}}+\frac{\partial \hat{f}}{\partial \vec{r}}\cdot \frac{\partial \hat{\varepsilon}}{\partial \vec{p}}\right ) \nonumber\\
&-&\frac{1}{2}\left ( \frac{\partial \hat{\varepsilon}}{\partial
\vec{r}}\cdot \frac{\partial \hat{f}}{\partial
\vec{p}}+\frac{\partial \hat{f}}{\partial \vec{p}}\cdot
\frac{\partial \hat{\varepsilon}}{\partial \vec{r}}\right )=0,
\end{eqnarray}
where $\hat{\varepsilon}$ and $\hat{f}$ are from the Wigner
transformation of the energy and phase-space density matrix,
respectively, and they can be decomposed into their scalar and
vector parts, i.e.,
\begin{eqnarray}
\hat{\varepsilon}(\vec{r},\vec{p})&=&\varepsilon(\vec{r},\vec{p})\hat{I}+\vec{h}(\vec{r},\vec{p})\cdot \vec{\sigma},\label{ener} \\
\hat{f}(\vec{r},\vec{p}) &=&
f_{0}(\vec{r},\vec{p})\hat{I}+\vec{g}(\vec{r},\vec{p})\cdot\vec{\sigma},
\label{dens}
\end{eqnarray} where
$\vec{\sigma}=(\sigma_{x},\sigma_{y},\sigma_{z})$ and $\hat{I}$ are
respectively the Pauli matrices and the $2\times2$ unit matrix,
$\varepsilon$ and $f_{0}$ are the scalar part of the effective
single-particle energy $\hat{\varepsilon}$ and phase-space density $\hat{f}$,
respectively, and $\vec{h}$ and $\vec{g}$ are the corresponding
vector distributions. Adding the Uehling-Uhlenbeck collision term,
the resulting spin-dependent BUU equation can be used to describe the spin-dependent
dynamics in various systems.

While the spin-dependent BUU equation can be taken as the starting point of investigating spin dynamics in HICs, a consistent derivation of the EOMs of nucleon test particles for simulations is still lacking.
The situation is quite different for electron spin transport in solid state physics.
To our best knowledge, the treatment of electron spin transport relevant to the present study mostly follows two approaches. One way (Method I) is to start from a model Hamiltonian and use the canonical EOMs for the time evolution of the electron's coordinate and momentum, while the time evolution of the electron's spin is given by its commutation relation with the model Hamiltonian as in the Heisenberg picture of quantum mechanics~\cite{Sin04,Sch06,Son10}. Moreover, an adiabatic approximation is often used so that the time evolution of spin can be solved first. Inserting the solution for spin evolution into the EOMs for coordinate and momentum then leads to the Berry curvature terms~\cite{Sun99,Xia10,Hua16}. Another method (Method II) frequently used is to linearize the spin-dependent BUU equation for spin-up and spin-down particles separately through the relaxation time approximation~\cite{Zha04,Mor15,Str89,Val93}. In nuclear physics, EOMs of nucleon test particles should be derived consistently from the BUU transport equation used to model HICs. It is well known that the spin-independent BV equation can be solved numerically by using the test-particle method~\cite{Won82,Ber84}. In particular, it
was shown that the EOMs of test particles are identical to the canonical EOMs if only the lowest order term in expanding the Wigner function is
considered (see Ref.~\cite{Won03} and comments in Ref.~\cite{Kri07}). Applying the test-particle approach to solving the spin-dependent BUU equation for the first time, we found that the EOMs of nucleon test particles are similar to those for electrons obtained within Method I described above, albeit with different forms of interactions.

{\it Decomposition of the spin-dependent phase-space distribution
function and its evolution:}~To avoid confusion, we begin by first
commenting on the two approaches of deriving the BV equation with different
definitions of the spinor Wigner distribution function often used in the
literature. Equation~(\ref{BLE}) was derived by means of the density
matrix method and taking the semiclassical limit as outlined by
Smith and Jensen~\cite{Smi89}. It can be separated into two
equations governing the scalar and vector distributions,
respectively,
\begin{eqnarray}
\frac{\partial f_{0}}{\partial t}+\frac{\partial
\varepsilon}{\partial \vec{p}} \cdot \frac{\partial f_{0}}{\partial
\vec{r}} -\frac{\partial \varepsilon}{\partial \vec{r}} \cdot
\frac{\partial f_{0}}{\partial \vec{p}} + \frac{\partial
\vec{h}}{\partial \vec{p}} \cdot \frac{\partial \vec{g}}{\partial
\vec{r}}  - \frac{\partial \vec{h}}{\partial \vec{r}} \cdot
\frac{\partial \vec{g}}{\partial \vec{p}} = 0, \label{scal}
\end{eqnarray}
\begin{eqnarray}
\frac{\partial \vec{g}}{\partial t}&+&\frac{\partial \varepsilon}{\partial \vec{p}} \cdot \frac{\partial \vec{g}}{\partial \vec{r}} -\frac{\partial \varepsilon}{\partial \vec{r}} \cdot \frac{\partial \vec{g}}{\partial \vec{p}} + \frac{\partial f_{0}}{\partial \vec{r}} \cdot \frac{\partial \vec{h}}{\partial \vec{p}} \nonumber \\
&-& \frac{\partial f_{0}}{\partial \vec{p}} \cdot \frac{\partial \vec{h}}{\partial \vec{r}} + \frac{2\vec{g}\times\vec{h}}{\hbar} = 0. \label{vect}
\end{eqnarray}
Another way to get the spin-dependent BV equation is to start from
the time-dependent Hartree-Fock equations for the one-body density
matrix with spin degree of freedom and rewrite the equations with
the help of the Wigner transformation, see, e.g.,
Refs.~\cite{Ring80,Balb13}. In this way, one will get four coupled
equations which describe the time evolution of the four-component
Wigner phase-space densities from the $2\times2$ density matrix with
spin. The definition of the Wigner function of particles with
spin-$1/2$ was suggested in Refs.~\cite{Carr83,Conn84} as
\begin{eqnarray}
f_{\sigma,{\sigma }'}(\vec{r}\vec{p},t) &=& \int d^{3}s e^{-i\vec{p}\cdot \vec{s}/\hbar}\psi _{{\sigma }'}^{*}(\vec{r}-\frac{\vec{s}}{2},t)\psi _{\sigma  }(\vec{r}+\frac{\vec{s}}{2},t), \label{fsig} \\
f(\vec{r}\vec{p},t,0) &=& f_{1,1}(\vec{r}\vec{p},t)+f_{-1,-1}(\vec{r}\vec{p},t), \label{f0} \\
\tau(\vec{r}\vec{p},t,x) &=&
f_{-1,1}(\vec{r}\vec{p},t)+f_{1,-1}(\vec{r}\vec{p},t),
\label{fx} \\
\tau(\vec{r}\vec{p},t,y) &=&-i[f_{-1,1}(\vec{r}\vec{p},t)-f_{ 1,-1}(\vec{r}\vec{p},t)], \label{fy}\\
\tau(\vec{r}\vec{p},t,z) &=&
f_{1,1}(\vec{r}\vec{p},t)-f_{-1,-1}(\vec{r}\vec{p},t), \label{fz}
\end{eqnarray}
with $\sigma({\sigma}')=1$ for spin up and $-1$ for spin down. The above definitions are convenient in treating the expectation values of the spin components. Equation~(\ref{fsig}) gives the matrix components of the
Wigner function with spin degree of freedom. $f(\vec{r}\vec{p},t,0)$
is the ordinary Wigner phase-space density irrespective of the particle
spin, while $\tau(\vec{r}\vec{p},t,x)$, $\tau(\vec{r}\vec{p},t,y)$,
and $\tau(\vec{r}\vec{p},t,z)$, representing the three components of
the spin Wigner density $\vec{\tau}(\vec{r},\vec{p},t)$, are the
probabilities of the spin projection on the $x$, $y$, and $z$
directions, respectively. With the above definitions, the Wigner
density $f(\vec{r},\vec{p},t)$ in Eq.~(\ref{f0}) and the spin Wigner
density $\vec{\tau}(\vec{r},\vec{p},t)$ (Eqs.~(\ref{fx} - \ref{fz}))
can be expressed in terms of the $f_{0}(\vec{r},\vec{p},t)$ and
$\vec{g}(\vec{r},\vec{p},t)$ in Eq.~(\ref{dens}) as~\cite{Carr83}
\begin{eqnarray}
f(\vec{r},\vec{p},t) &=&  2 f_{0}(\vec{r},\vec{p},t), \label{f2} \\
\vec{\tau}(\vec{r},\vec{p},t) &=& 2 \vec{g}(\vec{r},\vec{p},t).
\label{g2}
\end{eqnarray}
In this way, the two approaches using two different definitions of the spinor Wigner function lead to exactly the same spin-dependent BV equation.

{\it Single-particle energy with spin-orbit interaction:} While our derivation is general, to be specific, for the
spin-dependent part of the single-particle Hamiltonian in Eq.~(\ref{ener}) we take the Skyrme-type
effective two-body interaction including the spin-orbit coupling~\cite{Vau72,Eng75}
\begin{eqnarray}
\hat{h}_{q}^{so}&=&-\frac{1}{2}W_{0}\nabla\cdot (\vec{J}+\vec{J_{q}})+\vec{\sigma}\cdot [-\frac{1}{2}W_{0}\nabla \times (\vec{j}+\vec{j_{q}})] \nonumber \\
&+&\frac{1}{4i}W_{0}[(\nabla\times \vec{\sigma})\cdot
\nabla(\rho+\rho_{q})+\nabla(\rho+\rho_{q})\cdot (\nabla\times \vec{\sigma})] \nonumber \\
&-&\frac{1}{4i}W_{0}[\nabla\cdot (\nabla\times
(\vec{s}+\vec{s_{q}}))+ (\nabla\times (\vec{s}+\vec{s_{q}}))\cdot
\nabla], \label{hso}
\end{eqnarray}
where $q = n$ or $p$ is the isospin index, and $\rho$, $\vec{s}$,
$\vec{j}$, and $\vec{J}$ are the number, spin, momentum, and
spin-current densities, respectively. According to the definition of
the Wigner function in Eq.~(\ref{fsig}), these densities can be
directly expressed as~\cite{Vau72,Eng75}
\begin{eqnarray}
\rho(\vec{r}) &=& \int d^{3}p f(\vec{r},\vec{p}), \label{rhor}\\
\vec{s}(\vec{r}) &=& \int d^{3}p \vec{\tau}(\vec{r},\vec{p}),  \\
\vec{j}(\vec{r}) &=& \int d^{3}p \frac{\vec{p}}{\hbar}f(\vec{r},\vec{p}),  \\
\vec{J}(\vec{r}) &=& \int d^{3}p \frac{\vec{p}}{\hbar} \times
\vec{\tau}(\vec{r},\vec{p}). \label{Jr}
\end{eqnarray}
After a Wigner transformation, the Eq.~(\ref{hso}) can be readily
expressed in terms of the above densities as
\begin{equation}
h_{q}^{so}(\vec{r},\vec{p}) = h_{1} +h_{4} + (\vec{h}_{2} +\vec{h}_{3} )\cdot \vec{\sigma} \label{hqso}
\end{equation}
 with $h_{1}, \vec{h}_{2}, \vec{h}_{3} $,
and $h_{4}$ given by
\begin{eqnarray}
h_{1} &=& -\frac{W_{0}}{2} \nabla_{\vec{r}} \cdot[\vec{J}(\vec{r})+\vec{J}_{q}(\vec{r})],    \label{h1} \\
\vec{h}_{2} &=&   -\frac{W_{0}}{2} \nabla_{\vec{r}} \times [\vec{j}(\vec{r})+\vec{j}_{q}(\vec{r})],     \\
\vec{h}_{3} &=&  \frac{W_{0}}{2} \nabla_{\vec{r}}[\rho(\vec{r})+\rho_{q}(\vec{r})]\times \vec{p} ,  \\
h_{4} &=& -\frac{W_{0}}{2} \nabla_{\vec{r}} \times
[\vec{s}(\vec{r})+\vec{s}_{q}(\vec{r})]\cdot \vec{p}.    \label{h4}
\end{eqnarray}
Comparing with Eq.~(\ref{ener}), the effective single-particle
energy $\hat{\varepsilon}$ can be written as
\begin{eqnarray}
\varepsilon_{q}(\vec{r},\vec{p}) &=& \frac{p^2}{2m} +U_{q}+h_{1}+h_{4}, \label{epsilon} \\
\vec{h}_{q}(\vec{r},\vec{p}) &=& \vec{h}_{2} +\vec{h}_{3},\label{hqve}
\end{eqnarray}
where $U_{q}$ is the spin-independent mean-field potential. The
nuclear tensor force can be implemented in a similar way if needed,
once the scalar and the vector components are decomposed from the
corresponding energy-density functional.

{\it Spin-dependent EOMs of test particles:}~We now derive the EOMs
from the decoupled spin-dependent BV equation [Eqs.~(\ref{scal}) and
(\ref{vect})] by using the test-particle method~\cite{Won82,Won03}.
The vector part $\vec{g}(\vec{r},\vec{p})$ of the spinor Wigner
function distribution in Eq.~(\ref{dens}) can be represented by a
real unit vector $\vec{n}$ times a scalar function
$f_{1}(\vec{r},\vec{p})$, i.e.,
\begin{equation}
\vec{g}(\vec{r},\vec{p}) =\vec{n}f_{1}(\vec{r},\vec{p}).
\label{nvec}
\end{equation}
Here we assume that $\vec{n}$ is
independent of $\vec{r}$ and $\vec{p}$, which is valid if
$\vec{n}$ evolves much faster than the phase-space coordinates $\vec{r}$ and $\vec{p}$ or if $\vec{n}$ is a global constant.
Under this assumption and by substituting Eq.~(\ref{nvec}) into
Eqs.~(\ref{scal}) and (\ref{vect}), we obtain
\begin{eqnarray}
\frac{\partial f_{0}}{\partial t}&+&\frac{\partial \varepsilon}{\partial \vec{p}} \cdot \frac{\partial f_{0}}{\partial \vec{r}} -\frac{\partial \varepsilon}{\partial \vec{r}} \cdot \frac{\partial f_{0}}{\partial \vec{p}} + (\frac{\partial \vec{h}}{\partial \vec{p}} \cdot \vec{n}) \cdot \frac{\partial f_{1}}{\partial \vec{r}} \nonumber \\
&-& (\frac{\partial \vec{h}}{\partial \vec{r}} \cdot \vec{n}) \cdot  \frac{\partial f_{1}}{\partial \vec{p}} \approx 0, \label{scal1} \\
\frac{\partial f_{1}}{\partial t} \vec{n} &+&(\frac{\partial \varepsilon}{\partial \vec{p}} \cdot \frac{\partial f_{1}}{\partial \vec{r}})\vec{n} -(\frac{\partial \varepsilon}{\partial \vec{r}} \cdot \frac{\partial f_{1}}{\partial \vec{p}})\vec{n} + \frac{\partial f_{0}}{\partial \vec{r}} \cdot \frac{\partial \vec{h}}{\partial \vec{p}} \nonumber \\
&-& \frac{\partial f_{0}}{\partial \vec{p}} \cdot \frac{\partial
\vec{h}}{\partial \vec{r}} +
(\frac{2\vec{n}\times\vec{h}}{\hbar}+\frac{\partial
\vec{n}}{\partial t})f_{1} \approx 0. \label{vect1}
\end{eqnarray}
Generally, the magnitude of the Poisson bracket $\{f_0,\vec{h}\}$, i.e., $(\partial
f_{0}/\partial \vec{r}) \cdot (\partial \vec{h}/\partial
\vec{p})-(\partial f_{0}/\partial \vec{p}) \cdot (\partial
\vec{h}/\partial \vec{r})$, is much smaller than that of $\vec{h}$ or
$\{f_{0},\epsilon\}$. In this approximation and by separating
components parallel and perpendicular to $\vec{n}$,
Eq.~(\ref{vect1}) can be divided into two parts, i.e.,
\begin{eqnarray}
\frac{\partial f_{1}}{\partial t} \vec{n} &+&(\frac{\partial \varepsilon}{\partial \vec{p}} \cdot \frac{\partial f_{1}}{\partial \vec{r}})\vec{n} -(\frac{\partial \varepsilon}{\partial \vec{r}} \cdot \frac{\partial f_{1}}{\partial \vec{p}})\vec{n} + \frac{\partial f_{0}}{\partial \vec{r}} \cdot \frac{\partial \vec{h}}{\partial \vec{p}} \nonumber \\
&-& \frac{\partial f_{0}}{\partial \vec{p}} \cdot \frac{\partial \vec{h}}{\partial \vec{r}} =0, \label{vect2} \\
\frac{\partial \vec{n}}{\partial t} &\approx&
\frac{2\vec{h}\times\vec{n}}{\hbar}. \label{nte}
\end{eqnarray}
As $\vec{n}$ is a unit vector, we have $\vec{n}\cdot \vec{n} =1$. By
taking the inner product of $\vec{n}$ with Eq.~(\ref{vect2}) (or Eq.~(\ref{vect1}))
on both sides, one obtains
\begin{eqnarray}
\frac{\partial f_{1}}{\partial t} &+&\frac{\partial \varepsilon}{\partial \vec{p}} \cdot \frac{\partial f_{1}}{\partial \vec{r}} -\frac{\partial \varepsilon}{\partial \vec{r}} \cdot \frac{\partial f_{1}}{\partial \vec{p}} + \frac{\partial f_{0}}{\partial \vec{r}} \cdot (\frac{\partial \vec{h}}{\partial \vec{p}}\cdot \vec{n}) \nonumber \\
&-& \frac{\partial f_{0}}{\partial \vec{p}} \cdot (\frac{\partial \vec{h}}{\partial \vec{r}}\cdot \vec{n}) = 0. \label{vect3}
\end{eqnarray}
Adding and subtracting Eqs.~(\ref{scal1}) and (\ref{vect3}), we get
two equations for two types of particles with phase-space distribution
functions $f^{\pm}=f_0 \pm f_1$, i.e.,
\begin{eqnarray}
\frac{\partial f^+}{\partial t} + \left(\frac{\partial
\epsilon}{\partial \vec{p}} + \frac{\partial V_{hn}}{\partial
\vec{p}}\right) \cdot \frac{\partial f^+}{\partial \vec{r}} -
\left(\frac{\partial \epsilon}{\partial \vec{r}} + \frac{\partial
V_{hn}}{\partial \vec{r}}\right) \cdot \frac{\partial f^+}{\partial
\vec{p}} &=& 0, \label{f+}\\
\frac{\partial f^-}{\partial t} + \left(\frac{\partial
\epsilon}{\partial \vec{p}} - \frac{\partial V_{hn}}{\partial
\vec{p}}\right) \cdot \frac{\partial f^-}{\partial \vec{r}} -
\left(\frac{\partial \epsilon}{\partial \vec{r}} - \frac{\partial
V_{hn}}{\partial \vec{r}}\right) \cdot \frac{\partial f^-}{\partial
\vec{p}} &=& 0,\label{f-}
\end{eqnarray}
with $V_{hn}\equiv\vec{h} \cdot \vec{n}$. It can be understood from
Eqs.~(\ref{dens}) and (\ref{nvec}) that $f^+$ and $f^-$ are the
eigenfunctions of $\hat{f}$, representing the phase-space
distributions of particles with their spin in $+\vec{n}$ and
$-\vec{n}$ directions, respectively, i.e., spin-up and spin-down
particles.

Following the test-particle method and using an auxiliary variable
$\vec{s}$, the time evolution of the Wigner function
$f^{\pm}(\vec{r},\vec{p})$ can be expressed as~\cite{Won03}
\begin{eqnarray}
f^{\pm}(\vec{r},\vec{p},t) &=& \int \frac{d^{3}r_{0}d^{3}p_{0}d^{3}s}{(2\pi\hbar)^{3}} \exp \{i\vec{s}\cdot[\vec{p}-\vec{P}(\vec{r}_{0}\vec{p}_{0}\vec{s},t)]/\hbar\}  \nonumber \\
&\times& \delta
[\vec{r}-\vec{R}(\vec{r}_{0}\vec{p}_{0}\vec{s},t)]f^{\pm}(\vec{r}_{0},\vec{p}_{0},t_{0}),
\label{f}
\end{eqnarray}
where $f^{\pm}(\vec{r}_{0},\vec{p}_{0},t_{0})$ is the Wigner
functions at time $t_{0}$ with the initial conditions
$\vec{R}(\vec{r}_{0}\vec{p}_{0}\vec{s},t_{0}) = \vec{r}_{0}$ and
$\vec{P}(\vec{r}_{0}\vec{p}_{0}\vec{s},t_{0}) = \vec{p}_{0}$. Our
main task is now to find the new phase-space coordinates
$\vec{R}(\vec{r}_{0}\vec{p}_{0}\vec{s},t)$ and
$\vec{P}(\vec{r}_{0}\vec{p}_{0}\vec{s},t)$ and to obtain the Wigner
function at the next time step $t=t_{0}+\Delta{t}$ with a small
increment $\Delta{t}$. By substituting Eq.~(\ref{f}) into
Eqs.~(\ref{f+}) and (\ref{f-}), we obtain
\begin{eqnarray}
&[-&\frac{\partial \vec{R}(\vec{r}_{0}\vec{p}_{0}\vec{s},t)}{\partial t}+\frac{\partial \varepsilon}{\partial \vec{p}}]\cdot\frac{\partial f^{\pm}(\vec{r},\vec{p},t)}{\partial \vec{r}} \pm\frac{\partial V_{hn}}{\partial \vec{p}} \cdot \frac{\partial f^{\pm}(\vec{r},\vec{p},t)}{\partial\vec{r}} \nonumber \\
&+& \int \frac{d^{3}r_{0}d^{3}p_{0}d^{3}s}{(2\pi\hbar)^{3}} \{f^{\pm}(\vec{r}_{0},\vec{p}_{0},t_{0})[\frac{-i\vec{s}}{\hbar}\cdot \frac{\partial \vec{P}(\vec{r}_{0}\vec{p}_{0}\vec{s},t)}{\partial t} \nonumber \\
&-&\frac{[\varepsilon(\vec{r}-\frac{\vec{s}}{2},t)-\varepsilon(\vec{r}
+\frac{\vec{s}}{2},t)]}{i\hbar}] \nonumber \\
&\mp&
f^{\pm}(\vec{r}_{0},\vec{p}_{0},t_{0})[\frac{[V_{hn}(\vec{r}-\frac{\vec{s}}{2},t)-V_{hn}(\vec{r}
+\frac{\vec{s}}{2},t)]}{i\hbar}]\} \nonumber \\
&\times&\exp\{i\vec{s}\cdot[\vec{p}-\vec{P}(\vec{r}_{0}\vec{p}_{0}\vec{s},t)]/\hbar\} \nonumber \\
&\times&\delta [\vec{r}-\vec{R}(\vec{r}_{0}\vec{p}_{0}\vec{s},t)]
=0. \label{ft}
\end{eqnarray}
Comparing similar terms in the above equation, we get the following
equations for $\vec{R}$ and $\vec{P}$, respectively, i.e.,
\begin{eqnarray}
\left[-\frac{\partial
\vec{R}(\vec{r}_{0}\vec{p}_{0}\vec{s},t)}{\partial t}+\frac{\partial
\varepsilon}{\partial \vec{p}}\right]\cdot\frac{\partial
f^{\pm}(\vec{r},\vec{p},t)}{\partial \vec{r}}
&\pm& \frac{\partial V_{hn}}{\partial \vec{p}} \cdot \frac{\partial f^{\pm}(\vec{r},\vec{p},t)}{\partial\vec{r}}  \nonumber \\
&=& 0, \label{r1}
\end{eqnarray}
and
\begin{eqnarray}
&& f^{\pm}(\vec{r}_{0},\vec{p}_{0},t_{0})\{\frac{-i\vec{s}}{\hbar}\cdot \frac{\partial \vec{P}(\vec{r}_{0}\vec{p}_{0}\vec{s},t)}{\partial t} \nonumber \\
&&-\frac{[\varepsilon(\vec{r}-\frac{\vec{s}}{2},t)-\varepsilon(\vec{r}
+\frac{\vec{s}}{2},t)]}{i\hbar}\}\mp f^{\pm}(\vec{r}_{0},\vec{p}_{0},t_{0}) \nonumber \\
&&\times
\{\frac{[V_{hn}(\vec{r}-\frac{\vec{s}}{2},t)-V_{hn}(\vec{r}+\frac{\vec{s}}{2},t)]}{i\hbar}\}=0.
\label{p1}
\end{eqnarray}
In order to satisfy the above two equations with arbitrary
$f^{\pm}(\vec{r},\vec{p},t)$ and
$f^{\pm}(\vec{r}_{0},\vec{p}_{0},t_{0})$, we get the equations of
motion for $\vec{R}$
\begin{equation}
\frac{\partial \vec{R}(\vec{r}_{0}\vec{p}_{0}\vec{s},t)}{\partial t}
= \frac{\partial \varepsilon}{\partial \vec{p}} \pm \frac{\partial
V_{hn}}{\partial\vec{p}}, \label{rf}
\end{equation}
and for $\vec{P}$
\begin{eqnarray}
\vec{s}\cdot \frac{\partial \vec{P}(\vec{r}_{0}\vec{p}_{0}\vec{s},t)}{\partial t} = [\varepsilon(\vec{r}-\frac{\vec{s}}{2},t)-\varepsilon(\vec{r}
+\frac{\vec{s}}{2},t)]  \nonumber \\
\pm
[V_{hn}(\vec{r}-\frac{\vec{s}}{2},t)-V_{hn}(\vec{r}+\frac{\vec{s}}{2},t)].
\label{ps}
\end{eqnarray}
Expanding Eq.~(\ref{ps}) in $\vec{s}$ and keeping only the lowest
order term, we then obtain
\begin{equation}
\frac{\partial \vec{P}(\vec{r}_{0}\vec{p}_{0}\vec{s},t)}{\partial t} = -\frac{\partial
 \varepsilon}{\partial \vec{r}} \mp \frac{\partial V_{hn}}{\partial \vec{r}}. \label{pf}
\end{equation}
Considering Eqs.~(\ref{hqso}-\ref{hqve}) and combining
Eqs.~(\ref{rf}), (\ref{pf}), and (\ref{nte}), the EOMs of test
particles for solving the spin-dependent BV equation with the
spin-orbit interaction are thus
\begin{eqnarray}
\frac{\partial \vec{R}}{\partial t} &=& \frac{\vec{p}}{m}+\nabla_{\vec{p}} (h_{1}+h_{4})\pm\nabla_{\vec{p}}(\vec{h}_{2}\cdot \vec{n}+\vec{h}_{3}\cdot \vec{n}), \label{rmt}\\
\frac{\partial \vec{P}}{\partial t} &=& -\nabla_{\vec{r}}U_{q}-\nabla_{\vec{r}}(h_{1}+h_{4})
\mp\nabla_{\vec{r}}(\vec{h}_{2}\cdot \vec{n}+\vec{h}_{3}\cdot \vec{n}), \label{pmt}\\
\frac{\partial \vec{n}}{\partial t} &=& \frac{2(\vec{h}_{2}+\vec{h}_{3})\times\vec{n}}{\hbar}, \label{nmt}
\end{eqnarray}
with the upper sign for $f^+$ and lower sign for $f^-$,
respectively, and $h_{1}$, $\vec{h}_{2}$, $\vec{h}_{3}$, and $h_{4}$
given by Eqs.~(\ref{h1}-\ref{h4}). According to the definition of
the spinor Wigner phase-space density distribution in
Eqs.~(\ref{dens}) and (\ref{nvec}), $\vec{n}$ is the direction (unit
vector) of the local spin polarization in 3-dimensional coordinate
space, which can be expressed by the test-particle
method~\cite{Won82,Ber84} as
\begin{equation}
\vec{n} = \frac{\sum_i \vec{n}_i
\delta(\vec{r}-\vec{r}_i)\delta(\vec{p}-\vec{p}_i)}{|\sum_i
\vec{n}_i \delta(\vec{r}-\vec{r}_i)\delta(\vec{p}-\vec{p}_i)|},
\end{equation}
with $\vec{n}_{i}$ being the spin expectation direction of the $i$th nucleon.
Based on Eqs.~(\ref{f0}-\ref{fz}), the scalar Wigner density
distribution $f(\vec{r},\vec{p})$ and the vector Wigner density
distribution $\vec{\tau}(\vec{r},\vec{p})$ can be expressed by the
test-particle method as
\begin{eqnarray}
f(\vec{r},\vec{p}) &=&  \frac{1}{N_{TP}}\sum_{\rm i} \delta(\vec{r}-\vec{r}_i)\delta(\vec{p}-\vec{p}_i), \\
\vec{\tau}(\vec{r},\vec{p}) &=& \frac{1}{N_{TP}}\sum_{\rm i}
\vec{n_i} \delta(\vec{r}-\vec{r}_i)\delta(\vec{p}-\vec{p}_i),
\label{tauden}
\end{eqnarray}
with $N_{TP}$ being the number of test particles per nucleon.
In this way, the number, spin, momentum, and spin-current densities
can also be calculated via
\begin{eqnarray}
\rho(\vec{r}) &=& \frac{1}{N_{TP}}\sum_{\rm i} \delta(\vec{r}-\vec{r}_i),\\
\vec{s}(\vec{r}) &=& \frac{1}{N_{TP}}\sum_{\rm i} \vec{n_i} \delta(\vec{r}-\vec{r}_i),\label{sden}\\
\vec{j}(\vec{r}) &=& \frac{1}{N_{TP}}\sum_{\rm i} \frac{\vec{p}_i}{\hbar} \delta(\vec{r}-\vec{r}_i),\\
\vec{J}(\vec{r}) &=& \frac{1}{N_{TP}}\sum_{\rm i}
(\frac{\vec{p}_i}{\hbar} \times \vec{n_i})
\delta(\vec{r}-\vec{r}_i).\label{Jden}
\end{eqnarray}

{\it Quantum nature of spin:}~The above EOMs for the phase-space coordinates are the same as
the canonical equations in Hamiltonian dynamics, and the EOM of spin is the same as that in
the Heisenburg picture of quantum mechanics. These EOMs have been
applied in our previous studies~\cite{Xu13,Xia14,Xu15}. As first demonstrated by the Stern-Gerlach experiment~\cite{ST22},
the projection of spin onto any reference direction used in measurements is quantized.
In non-central HICs, the angular momentum is in the $y$ direction perpendicular to the
reaction plane ($x$-$o$-$z$ plane). It is thus natural to set $y$
direction as the third (magnetic) spin direction. We then fix
$\vec{n}=\hat{y}$ in Eqs.~(\ref{rmt}) and (\ref{pmt}) and set
$\vec{n}_i= \pm \hat{y}$ in Eqs.~(\ref{tauden}), (\ref{sden}), and
(\ref{Jden}) depending on whether the $i$th particle is spin-up or
spin-down with respect to the $y$ axis. In this way the time evolution of $\vec{n}$ (Eq.~(\ref{nmt})) is not
needed.  Thus, as describing the isospin dynamics with separate EOMs for neutrons and protons \cite{LCK08},
we now have separate EOMs for spin-up (upper sign) and spin-down (lower sign) particles
\begin{eqnarray}
\frac{\partial \vec{R}}{\partial t} &=& \frac{\vec{p}}{m}+\nabla_{\vec{p}} (h_{1}+h_{4})\pm\nabla_{\vec{p}}(h_{2y}+h_{3y}), \label{qrmt}\\
\frac{\partial \vec{P}}{\partial t} &=& -\nabla_{\vec{r}}U_{q}-\nabla_{\vec{r}}(h_{1}+h_{4})
\mp\nabla_{\vec{r}}(h_{2y}+h_{3y}). \label{qpmt}
\end{eqnarray}
It is seen that the $h_{2y}\equiv \vec{h}_2\cdot \vec{n}$ and $h_{3y}\equiv \vec{h}_3\cdot \vec{n}$ lead to the spin-dependent motion while the $h_1$ and $h_4$
affect the global motion in phase space.

To summarize, the spin-dependent Boltzmann-Vlasov equation can be
solved by extending the test-particle method. Considering the quantum nature of spin and choosing the direction of total angular momentum in heavy-ion reactions as a reference of measuring nucleon spin,
the EOMs of spin-up and spin-down nucleons are derived. The key elements affecting the spin dynamics in heavy-ion collisions are identified.
The derived EOMs of test particles provide the theoretical foundation of simulating spin-dependent dynamics in intermediate-energy heavy-ion collisions.
Future comparisons of model simulations with experimental data will help constrain the poorly known in-medium nucleon spin-orbit coupling.

This work was supported by the Major State Basic Research
Development Program (973 Program) of China under Contract Nos.
2015CB856904 and 2014CB845401, the National Natural Science
Foundation of China under Grant Nos. 11320101004, 11475243, and
11421505, the "100-talent plan" of Shanghai Institute of Applied
Physics under Grant Nos. Y290061011 and Y526011011 from the Chinese
Academy of Sciences, the Shanghai Key Laboratory of Particle Physics
and Cosmology under Grant No. 15DZ2272100, the "Shanghai Pujiang
Program" under Grant No. 13PJ1410600, the US National Science
Foundation under Grant No. PHY-1068022, the U.S. Department of
Energy, Office of Science, under Award Number de-sc0013702, and the
CUSTIPEN (China-U.S. Theory Institute for Physics with Exotic
Nuclei) under the US Department of Energy Grant No. DEFG02-
13ER42025.


\begin{thebibliography}{99}

\bibitem{May49} M. G. Mayer, Phys. Rev. {\bf75}, 1969 (1949).

\bibitem{May55} M. G. Mayer and J. H. D. Jenson, {\it Elementary Theory of Nuclear Shell Structure} (Wiley, New York, 1955).

\bibitem{Lal98} G. A. Lalazissis {\it et al.}, Phys. Lett. B {\bf 418}, 7 (1998).

\bibitem{Mor08} M. Morjean {\it et al.}, Phys. Rev. Lett. {\bf 101}, 072701 (2008).

\bibitem{Sch04} J. P. Schiffer {\it et al.}, Phys. Rev. Lett. {\bf 92}, 162501 (2004).

\bibitem{Ben09} M. Bender {\it et al.}, Phys. Rev. C {\bf80}, 064302 (2009).

\bibitem{Che95} B. Chen {\it et al.}, Phys. Lett. B {\bf355}, 37 (1995).

\bibitem{Kra07} K.-L. Kratz, K. Farouqi, and B. Pfeiffer, Prog.
Part. Nucl. Phys. \textbf{59}, 147 (2007).

\bibitem{Uma86} A. S. Umar {\it et al.}, Phys. Rev. Lett. {\bf56}, 2793 (1986).

\bibitem{Ich12} Y. Ichikawa {\it et al.}, Nature Phys. {\bf8}, 918 (2012).

\bibitem{Sch94} W. D. Schmidt-Ott {\it et al.}, Z. Phys. A \textbf{350}, 215
(1994).

\bibitem{Gro03} D. E. Groh {\it et al.}, Phys. Rev. Lett. {\bf90}, 202502 (2003).

\bibitem{Tur06} K. Turzo {\it et al.}, Phys. Rev. C {\bf73}, 044313 (2006).

\bibitem{Xu13} J. Xu and B. A. Li, Phys. Lett. B {\bf724}, 346 (2013).

\bibitem{Xia14} Y. Xia, J. Xu, B. A. Li, and W. Q. Shen, Phys. Rev. C {\bf89}, 064606 (2014).

\bibitem{Xu15} J. Xu, B. A. Li, W. Q. Shen, and Y. Xia, Front. Phys. {\bf 10}, 102501 (2015).

\bibitem{RHICspin} \url{https://wiki.bnl.gov/rhicspin/Presentations}

\bibitem{NSAC} \url{http://science.energy.gov/~/media/np/nsac/pdf/2015LRP/2015_LRPNS_091815.pdf}

\bibitem{Won03} C. Y. Wong, J. Opt. B {\bf5}, S420 (2003).

\bibitem{Smi89} H. Smith and H. H. Jensen, {\it Transport Phenomena} (Oxford University press, Oxford, 1989).

\bibitem{Hal69} B. I. Halperin and P. C. Hohenberg, Phys. Rev. {\bf188}, 898 (1969).

\bibitem{Joh84} B. R. Johnson {\it et al.}, Phys. Rev. Lett. {\bf52}, 1508 (1984).

\bibitem{Ber96} L. Berger, Phys. Rev. B {\bf54}, 9353 (1996).

\bibitem{Sch06} J. Schlimann, Int. J. Mod. Phys. B {\bf20}, 1015 (2006).

\bibitem{Mor15} K. Morawetz, Phys. Rev. B {\bf92}, 245425 (2015).

\bibitem{Sin15} J. Sinova {\it et al.}, Rev. Mod. Phys. {\bf87}, 1213 (2015).


\bibitem{Conn84} R. F. O'Connell and E. P. Wigner, Phys. Rev. A {\bf30}, 2613 (1984).

\bibitem{Ring80} P. Ring and P. Schuck, {\it The Nuclear Many-Body Problem} (Springer, Berlin, 1980).

\bibitem{Balb13} E. B. Balbutsev, I. V. Molodtsova, and P. Schuck, Phys. Rev. C {\bf88}, 014306 (2013).

\bibitem{Sin04} J. Sinova {\it et al.}, Phys. Rev. Lett. \textbf{92}, 126603 (2004).

\bibitem{Son10} E. B. Sonin, Adv. Phys. {\bf59}, 181 (2010).

\bibitem{Sun99} G. Sundaram and Q. Niu, Phys. Rev. B \textbf{59}, 14915 (1999).

\bibitem{Xia10} D. Xiao, M. C. Chang, and Q. Niu, Rev. Mod. Phys. \textbf{82}, 1959 (2010).

\bibitem{Hua16} X. G. Huang, Sci. Rep. \textbf{6}, 20601 (2016).

\bibitem{Zha04} J. W. Zhang, P. Levy, S. F. Zhang, and V. Antropov, Phys. Rev. Lett. {\bf93}, 256602 (2004).

\bibitem{Str89} G. Strinati, C. Castellani, and C. Di Castro, Phys. Rev. B {\bf40}, 12237 (1989).

\bibitem{Val93} T. Valet and A. Fert, Phys. Rev. B {\bf48}, 7099 (1993).

\bibitem{Won82} C. Y. Wong, Phys. Rev. C {\bf25}, 1460 (1982).

\bibitem{Ber84} G. F. Bertsch, H. Kruse, and S. Das Gupta, Phys. Rev. C \textbf{29}, 673 (1984).

\bibitem{Kri07} M. I. Krivoruchenko, B. V. Martemyanov, and C.
Fuchs, Phys. Rev. C \textbf{76}, 059801 (2007).

\bibitem{Carr83} P. Carruthers and F. Zachariasen, Rev. Mod. Phys. {\bf55}, 245 (1983).

\bibitem{Vau72} D. Vautherin and D. M. Brink, Phys. Rev. C {\bf5}, 626 (1972).

\bibitem{Eng75} Y. M. Engel {\it et al.}, Nucl. Phys. A {\bf249}, 215 (1975).

\bibitem{ST22} W. Gerlach and O. Stern, Z. Phys. \textbf{8}, 110
(1922); Z. Phys. \textbf{9}, 349 (1922); Z. Phys. \textbf{9}, 353
(1922).


\bibitem{LCK08} B. A. Li, L. W. Chen, and C. M. Ko, Phys. Rep. \textbf{464}, 113 (2008).

\end{thebibliography}
\end{document}